\RequirePackage{fix-cm}
\documentclass{article}  
\usepackage[margin=1in]{geometry}
\usepackage[utf8]{inputenc}
\usepackage{natbib}
\usepackage{multirow}
\usepackage{appendix}
\usepackage{tikz}
\usepackage{graphicx}
\usepackage{comment}
\usepackage{hyperref}
\usepackage{color}
\usepackage{colortbl}
\usepackage{siunitx}
\usepackage{adjustbox}
\usepackage{csquotes}
\usepackage{csquotes}
\usepackage{graphicx}
\usepackage{algorithm}
\usepackage{algpseudocode}
\usepackage{mathtools}
\usepackage{amsfonts}
\usepackage{natbib}
\usepackage{booktabs}
\usepackage{tablefootnote}
\usepackage{threeparttable}
\usepackage{booktabs}
\usepackage{siunitx}
\usepackage{float}
\usepackage{tabularx}

\begin{document}


\title{Time Series Embedding and Combination of Forecasts: A Reinforcement Learning Approach}

\author{
Marcelo C. Medeiros\footnote{University of Illinois at Urbana-Champaign. Department of Economics, 214 David Kinley Hall. 
Urbana, IL 61801. Email: marcelom@illinois.edu.}
\and
Jeronymo M. Pinto\footnote{Esplanada dos Ministérios, F Block, Floor 1, Ministry of Economics - SIT, Brasília/DF, Brasil. Email: jeronymomp@gmail.com.} \footnote{ Corresponding author.}
}

\maketitle

\setcounter{page}{0}\thispagestyle{empty}

\maketitle

\begin{abstract}
The forecasting combination puzzle is a well-known phenomenon in forecasting literature, stressing the challenge of outperforming the simple average when aggregating forecasts from diverse methods. This study proposes a Reinforcement Learning - based framework as a dynamic model selection approach to address this puzzle. Our framework is evaluated through extensive forecasting exercises using simulated and real data. Specifically, we analyze the M4 Competition dataset and the Survey of Professional Forecasters (SPF). This research introduces an adaptable methodology for selecting and combining forecasts under uncertainty, offering a promising advancement in resolving the forecasting combination puzzle.
\end{abstract}





\def\spacingset#1{\renewcommand{\baselinestretch}%
{#1}\small\normalsize} \spacingset{1}

\spacingset{1.5}

\section{Introduction}
The study of forecast combinations has persistently captured considerable interest within the forecasting literature, particularly in Economics. A pivotal contribution was made by \cite{bates1969combination}, who explored the characteristics of forecast combinations and assessed whether integrating forecasts from multiple models could surpass the performance of simple averages. In an early review, \cite{clemen1989combining} concluded that equally weighting individual forecasts frequently constitutes the most efficacious approach. This conclusion led to the emergence of the ``\textbf{forecasting combination puzzle}''. Despite extensive empirical findings and numerous simulation studies, optimal forecast combination strategies often fail to surpass the simple average. This conundrum underscores a critical issue: no single model consistently prevails over others in all situations. Only if a universal ``winner'' could be determined for each scenario would a superior alternative to simple averaging emerge. 

Suppose that at a given moment $t$, and utilizing the available information set, an agent intends to construct a forecast $\widehat{Y}_{t+h|t}$ for a target variable $Y_{t+h}$, where $h>0$ denotes the forecast horizon. Furthermore, let us assume that at time $t$, the agent possesses $n$ possible forecasts derived from various methods or models, $\widehat{Y}_{t+h|t}^{(a)}$, $a=,\ldots,n$. Among these available models or methods, it is presumed that several canonical model combinations are incorporated. For instance, a simple average of forecasts might exemplify one such model. Consequently, at each period, the agent (forecaster) will take an action by selecting a forecast $\widehat{Y}_{t+h|t}^{(a_0)}$, $a_0\in\{1,\ldots,n\}$, based on a chosen criterion function. A common strategy would be to opt for the model with the lowest mean squared error (MSE) over a recent historical span, such as the last $\tau$ time intervals. Nevertheless, this approach may not prove optimal in contexts characterized by instability or structural breaks. Therefore, in this paper, we introduce an algorithm grounded in Reinforcement Learning (RL) designed to identify the most suitable model under analogous environmental conditions observed in antecedent periods. Our approach is loosely motivated by \citet{franklin2024news}, which presented a novel semantic search tool to identify historical news articles most similar to contemporary news queries.

\cite{feng2019reinforcement} and \cite{pinto2022machine} provided evidence of the potential of Reinforcement Learning (RL) for model selection and its adaptability to structural changes. However, no research definitively establishes RL as the optimal framework for forecasting in scenarios where the data-generating process (DGP) is unknown. In this paper, we propose a novel solution utilizing RL to dynamically and adaptively select the optimal weights, thereby enhancing forecasting accuracy. We validate our approach through a comprehensive set of experiments on simulated data as well as two real-world datasets: the M4 Competition dataset and the Survey of Professional Forecasters (SPF) dataset.\footnote{\url{https://forecasters.org/resources/time-series-data/}}\footnote{\url{https://www.philadelphiafed.org/}}

\section{Reinforcement Learning}

As described by \cite{elavarasan2020crop}, RL is a dynamic programming framework that rewards successful decisions. Unlike traditional Machine Learning (ML) approaches, RL agents learn through interaction with their environment and focus on optimizing actions based on cumulative rewards. This framework has garnered significant attention within the ML community; see, for example, \cite{ji2020spatio} and \cite{dong2020principled}.

Reinforcement Learning (RL) represents a Sequential Decision Problem (SDP), characterized by the following essential components.
\begin{itemize}
    \item \textbf{State Variable ($S_t$):} Encodes relevant information for agent's decision-making.
    \item \textbf{Decision Variables ($A_t$):} Actions taken by the agent, guided by a policy rule ($\pi$).
    \item \textbf{Exogenous Information ($\boldsymbol{W}_{t+1}$):} External factors (variables).
    \item \textbf{Transition Function:} Updates the state based on decisions and external information:
    \[
    S_{t+1} = S^M(S_t, A_t,\boldsymbol{W}_{t+1}),
    \]
    where $S^M$ is an arbitrary function that maps the previous state into the current state. The expected rewards for each state-action pair ($S_t,A_t$) are updated based on the environment's transition function and are continuously stored in the so-called \textbf{Q-table}, $Q(S_t,A_t)$.
    \item \textbf{Objective Function:} The goal is to maximize cumulative rewards or minimize penalties, $R_t$:
    \[
    v_\pi(S_t) = \mathbb{E}_\pi \left[\sum_{k=0}^\infty \gamma^k R_{t+k+1} \middle| S_t = s \right],
    \]
    where $\gamma$ is the discount factor.
\end{itemize}


Among reinforcement learning (RL) methods, \textbf{Temporal Difference (TD)} learning is widely used due to its ability to update estimates directly from experience, without requiring a model of the environment or waiting for the end of an episode — a crucial advantage in non-stationary settings \citep{sutton1988learning}. TD learning forms the foundation of value-based methods such as \textbf{Q-learning}, which maintains a Q-table containing estimates of the expected return for each state-action pair. The Q-table is updated incrementally using the Bellman equation:
\begin{equation}
    Q(S_t, A_t) \leftarrow Q(S_t, A_t) + \alpha \left[ R_{t+1} + \gamma \max_{a} Q(S_{t+1}, a) - Q(S_t, A_t) \right].
    \label{eq:td}
\end{equation}

In this formulation, $Q(S_t, A_t)$ is the current reward for taking action $A_t$ in state $S_t$, $\alpha$ is the learning rate ($0 < \alpha \leq 1$), $\gamma$ is the discount factor ($0 \leq \gamma < 1$), $R_{t+1}$ is the reward observed after the action, and $S_{t+1}$ is the next state. The term in brackets is the \textbf{TD error}, measuring the difference between the predicted and actual reward. The Q-value is adjusted in the direction that minimizes this error.

In our forecasting framework, we adapt this principle by interpreting each state $S_t$ as a snapshot of the cumulative Mean Squared Errors (MSEs) up to time $t$, and each action $A_t$ as the selection of a forecasting model. The reward $R_{t+1}$ corresponds to the negative forecasting error at time $t+1$, since our objective is to minimize error rather than maximize reward. 

This update enables the RL agent to learn which forecasting models perform best under varying historical error profiles, progressively enhancing model selection as more data becomes available. \textbf{The goal is to develop an algorithm capable of retaining past data patterns and selecting the most appropriate model for each scenario.} In this sense, the algorithm is designed to behave similarly to a researcher who, when faced with economic data, must determine the most suitable forecasting model for a given economic context.

\section{Research Design}

\subsection{State Embedding}

Consider that at time $t$, the forecaster assembles a $p\times t$ matrix $\boldsymbol{\mathcal{E}}_t$ comprising features of each model from the $n$ models/methods available, in conjunction with environmental factors. For instance, suppose the target variable is the next-period inflation. In this case, $\boldsymbol{\mathcal{E}}_t$ may include the squared errors for each period in time, the cumulative forecast errors, several other performance metrics of each model or method (mean squared or absolute errors, for instance), as well as macroeconomic variables to reflect the current state of the economy, word counts from news data, among many other features. It is noteworthy that $\boldsymbol{\mathcal{E}}_t$ may exhibit high dimensionality and encompass characteristics that facilitate the agent’s selection of the optimal model.

The objective is to encapsulate the information from $\boldsymbol{\mathcal{E}}_t$ into a low-dimensional vector of embeddings. This will be the state $\boldsymbol{S}_t$. This is achieved by initially performing a principal component analysis (PCA) on the matrix $\boldsymbol{\mathcal{E}}_t$ and selecting $k<p$ principal components. $\boldsymbol{S}_t$ will be the computed $k$ first principal components at time $t$. Note that, the values of $\boldsymbol{S}_j$, $j<t$, are not updated at this stage. Consequently, a table (Q-table) is constructed for each time period, wherein the columns comprise the embeddings and the rewards (for instance, the negative of the squared forecasting error for $Y_t$) for each forecasting model or method.

\subsection{Agent's Workflow}

The agent proceeds as follows. The first step is to compute the cosine similarity of the current embedding ($\boldsymbol{S}_t$) with all the past embeddings ($\boldsymbol{S}_j$, $j<t$) and select the period with the highest similarity. Call it $t_0$. If similarity exceeds a threshold $\eta$, set $Q(\boldsymbol{S}_t,a)=Q(\boldsymbol{S}_{t_0},a)$ for each action/model $a$, and the model with the highest reward at $t_0$ is selected to be the model to predict $Y_{t+h}$. Otherwise, a simple average of all models (or another benchmark model) is used.  Finally, based on the temporal difference equation, if the estimated reward differs from the observed value at $t+h$, the Q-table is updated accordingly. Otherwise, it remains unchanged. The Q-table is updated as
\[
Q(\boldsymbol{S}_t,a)\leftarrow Q(\boldsymbol{S}_t, a) + \alpha\left[ G_{t+h}(a) - Q(\boldsymbol{S}_t, a) \right],
\]
where $\alpha$ is the learning rate, and $G_t(a):=-(Y_{t+h}-\widehat{Y}^{(a)}_{t+h|t})^2$ is the observed reward from action (model) $a$. This approach iteratively refines model selection by learning from past error patterns, improving forecasting accuracy over time.

\section{Results}

In this section, we report empirical results of our proposed method. We adopt a simple version of the methodology described above, where $h=1$ and the matrix $\boldsymbol{\mathcal{E}}_t$ consists only of cumulative squared errors. All of these results, along with additional findings and code, are available in our open GitHub repository: \url{https://github.com/jeronymomp/reinforcement_learning_forecast}. For conciseness, we have chosen to present only the results described below.

\subsection{M4 Competition Dataset}

The M competition is an important forecasting competition \citep{bojer2021kaggle}. It assesses the accuracy of various methods across extensive sets of diverse series, encompassing a wide range of data domains and frequencies \citep{makridakis2022m5}. The M4 competition presented many methods to forecast a time series that were hard to beat. \footnote{All series and submitted point forecasts are extracted from the oficial competition webpage \url{https://github.com/Mcompetitions/M4-methods/tree/master} at 02/24/2024.}

According to \cite{makridakis2020m4}, the 100,000 time series of the M4 dataset come mainly from the Economics, Finance, Demographics and Industry areas. For the hourly data, we have 414 time series with different sizes of training sets that were used to forecast 48 observations. We used the forecasts from 61 models, including the submissions and some benchmarks proposed by the organizers. For more information about the competition, see \url{https://forecasters.org/resources/time-series-data/}. We opted to use the Hourly series from the M4 competition, since it is the longest one.

Our methodology allows selecting any of the 61 models submitted to the competition, including as well the simple average across all models. We tested our approach on all 414 competition experiments. While performance on individual series showed no single approach consistently excelling, this aligns with the competition's goal of identifying the method with the overall best performance through comprehensive ranking.

We computed the mean squared error (MSE) for all models, including RL, ranking the best-performing solutions. The average ranking, based on a model's position across experiments, remained consistent regardless of the chosen metric. Results for the MSE are summarized in Table \ref{tab:tabela10}.

\begin{table}[H]
\caption{M4 Competition: MSE} 
\centering
\begin{threeparttable}
\begin{tabular}{lr}
\toprule
Model &    MSE      \\
\midrule
\textbf{RL}                                     & \textbf{15.235}  \\
University of Oxford                            & 16.051  \\
ProLogistica Soft                               & 17.944  \\
Business Forecast Systems                       & 18.235  \\
University of A Coruña \& Monash University     & 18.387  \\
Individual - Nikzad                             & 18.866  \\
Individual - Kharaghani                         & 18.866  \\
Universiti Tun Hussein Onn Malaysia             & 19.636  \\
Uber Technologies                               & 19.700  \\
Individual - Jaganathan and Prakash             & 20.182  \\
\bottomrule
\end{tabular}
\label{tab:tabela10}
\end{threeparttable}
\end{table}

These results show that the RL method delivers superior performance, despite not being the top model in any experiment. Given the strong competition benchmarks, no model excelled universally. This underscores the RL method's adaptability in uncertain scenarios, selecting suitable forecasting models for varying environments.

\subsection{Survey of Professional Forecasters}

\subsubsection{The Survey}

The Survey of Professional Forecasters (SoPF), conducted quarterly by the Federal Reserve Bank of Philadelphia, is the oldest U.S. macroeconomic forecast survey.\footnote{\url{https://www.philadelphiafed.org/surveys-and-data/real-time-data-research/survey-of-professional-forecasters}.} It collects forecasts from diverse panelists across economic sectors for numerous variables. However, some data are missing, as not all experts submit forecasts consistently. Table \ref{tab:tabela11} lists all series.

\begin{table}[H]
\caption{Time Series with Experts Forecasts Published by the SoPF.}
\centering
\begin{threeparttable}
\begin{tabular}{lr}
\toprule
Series &    Code      \\
\midrule
Nominal Gross National Product/Gross Domestic Product	& NGPD \\
Price Index for Gross National Product/Gross Domestic Product 	& PGDP \\
Civilian Unemployment Rate  &	UNEMP \\
Industrial Production Index  &	INDPROD \\
Housing Starts  &	HOUSING \\
Real Gross National Product/Gross Domestic Product  &	RGDP \\
Real Personal Consumption Expenditures  &	RCONSUM \\
Real Nonresidential Fixed Investment  &	RNRESIN \\
Real Residential Fixed Investment  &	RRESINV \\
Government Consumption/Gross Investment  &	RFEDGOV \\
Real State/Local Government Consumption and Investment  &	RSLGOV \\
CPI Inflation Rate  &	CPI \\
PCE Inflation Rate  &	PCE \\
Core PCE Inflation Rate  &	COREPCE \\
Real Change in Private Inventories & RCBI \\
\bottomrule
\end{tabular}
\label{tab:tabela11}
\end{threeparttable}
\end{table}

Due to missing data, some time-series in Table \ref{tab:tabela11} were excluded due to missing data (EMP, RCONSUM, UNEMP). For the remaining series, we used the industry mean forecast as a benchmark, challenging RL to adapt without case-specific data. RL aims to identify the most suitable model, converging as at least the second-best choice. Results for 1-step-ahead forecasts are shown in Table \ref{tab:tabela12}, with MSE benchmarks.

\begin{table}[H]
\caption{Mean Squared Error for RL and the Mean Average of Industry's Experts.}
\centering
\begin{tabularx}{\textwidth}{lXXXXXX}
\toprule
 Series & Industry 1 & Industry 2 & Industry 3 & Simple Average & RL \\
\midrule
COREPCE  & 95.06   & 95.04   & 95.12   & 95.08   & 95.04   \\
CPI      & 230.81  & 230.44  & 230.53  & 230.59  & 230.52  \\
HOUSING  & 1333.92 & 1333.93 & 1333.91 & 1333.92 & 1333.92 \\
INDPROD  & 24.68   & 24.60   & 24.69   & 24.66   & 24.61   \\
NGDP     & 448.77  & 452.29  & 460.34  & 453.80  & 452.71  \\
PCE      & 95.94   & 95.69   & 95.77   & 95.80   & 95.78   \\
PGDP     & 30.14   & 30.17   & 30.10   & 30.14   & 30.10   \\
RCBI     & 40.84   & 42.14   & 42.33   & 41.64   & 41.63   \\
RCONSUM  & 2462.52 & 2463.00 & 2465.75 & 2463.76 & 2462.81 \\
RFEDGOV  & 364.78  & 364.46  & 362.87  & 364.04  & 362.87  \\
RGDP     & 2533.03 & 2554.39 & 2553.21 & 2546.88 & 2533.03 \\
RNRESIN  & 379.74  & 382.05  & 386.70  & 382.83  & 380.13  \\
RRESINV  & 317.68  & 318.44  & 317.82  & 317.98  & 318.08  \\
RSLGOV   & 734.91  & 735.26  & 734.18  & 734.78  & 734.47  \\
\bottomrule
\end{tabularx}
\label{tab:tabela12}
\end{table}

Table \ref{tab:tabela12} shows RL as the second-best option in several cases, such as INDPROD, PGDP, and RLSGOV, even outperforming all other methods in some instances. However, for NGDP and PCE, RL ranked third, falling short of a superior outcome.

The mean average of expert forecasts, a challenging benchmark, aligns with the forecast combination puzzle. Table \ref{tab:tabela13} provides the arithmetic average ranking across all experiments, indicating RL’s lower average ranking and confirming it as the best overall choice.

\begin{table}[H]
\caption{Arithmetic Average of the Ranking Obtained by the Simple Average of all Experts Forecasts and the RL method.}
\centering
\begin{threeparttable}
\begin{tabular}{lr}
\toprule
Forecast &    Average Ranking      \\
\midrule
Simple Average & 3.43 \\
Industry 1 & 3.32 \\
Industry 2 & 3.32 \\
Industry 3 & 3.00 \\
RL & \textbf{1.93} \\
\bottomrule
\end{tabular}
\label{tab:tabela13}
\end{threeparttable}
\end{table}


\section{Conclusion}

We developed a method to address the forecast combination puzzle using a RL framework. Testing on extensive experiments with real data, we found that RL effectively detects and switches to the most suitable model as needed, with better empirical results than the simple average. 

Our findings suggest a new forecasting approach for scenarios with limited information. Framing forecasting as a Sequential Decision Problem enables RL to learn from past data to optimize performance, surpassing traditional combination methods like the Simple Average, which lacks learning capabilities. This underscores the significance of learning as a novel approach to model selection, employing a legitimate artificial intelligence algorithm, as highlighted in \cite{silver2021reward}.

\textbf{Funding}

This research did not receive any specific grant from funding agencies in the public, commercial, or not-for-profit sectors.

\textbf{Declaration of competing interest}

The authors declare that they have no known competing financial
interests or personal relationships that could have appeared to influence
the work reported in this paper.

\textbf{Data availability}

All of these results, along with additional findings and simulation experiments conducted to test the method's robustness, are available in our open GitHub repository: \url{https://github.com/jeronymomp/reinforcement_learning_forecast}.

\clearpage
\bibliographystyle{plainnat}
\bibliography{sample.bib}

\end{document}